\documentclass[iop]{emulateapj}
\usepackage{apjfonts}
\usepackage{amssymb}
\usepackage{natbib}
\usepackage{appendix}

\def\ionii{{\it Ion2}}

\newcommand{\flyc}{\ifmmode  \mathrm{f}_\mathrm{esc}\mathrm{(LyC)} \else $\mathrm{f}_\mathrm{esc}\mathrm{(LyC)}$\fi}
\newcommand{\flya}{\ifmmode  \mathrm{f}_\mathrm{esc}\mathrm{(Ly\alpha)} \else $\mathrm{f}_\mathrm{esc}\mathrm{(Ly\alpha)}$\fi}
\newcommand{\dfel}{\ifmmode \Delta\log {\rm f}_{\rm EL} \else $\Delta\log$ f$_{\rm EL}$\fi}
\newcommand{\hst}{{\it HST}}

\newcommand{\hii}{\textrm{H}\textsc{ii}}

\newcommand{\oiii}{[\textrm{O}\textsc{iii}]}
\newcommand{\oii}{[\textrm{O}\textsc{ii}]}
\newcommand{\oiilam}{[\textrm{O}\textsc{ii}]\ensuremath{\lambda3727}}
\newcommand{\oiiiv}{[\textrm{O}\textsc{iii}]\ensuremath{\lambda5007}}

\newcommand{\oiiidoub}{[\textrm{O}~\textsc{iii}]\ensuremath{\lambda\lambda4959,5007}}
 
\newcommand{\ha}{\ifmmode {\rm H}\alpha \else H$\alpha$\fi}
\newcommand{\hb}{\ifmmode {\rm H}\beta \else H$\beta$\fi}
\newcommand{\lya}{\ifmmode {\rm Ly}\alpha \else Ly$\alpha$\fi}
\newcommand{\pg}{\ifmmode {\rm P}\gamma \else Pa$\gamma$\fi}
\newcommand{\lyb}{\ifmmode {\rm Ly}\beta \else Ly$\beta$\fi}
\newcommand{\lyg}{\ifmmode {\rm Ly}\gamma \else Ly$\gamma$\fi}
\newcommand{\ciii}{\textrm{C}\textsc{iii}]\ensuremath{\lambda1909}}

\newcommand{\nv}{\textrm{N}\textsc{v}\ensuremath{\lambda1240}}
\newcommand{\civ}{\textrm{C}\textsc{iv}\ensuremath{\lambda1550}}
\newcommand{\heii}{\textrm{He}\textsc{ii}\ensuremath{\lambda1640}}
\newcommand{\hei}{\textrm{He}\textsc{i}\ensuremath{\lambda10830}}

\def\ergs{\ifmmode \mathrm{erg\hspace{1mm}s}^{-1} \else erg s$^{-1}$\fi}

\def\micron{\ifmmode \mu\mathrm{m} \else $\mu$m\fi}
\def\msun{\ifmmode \mathrm{M}_{\odot} \else M$_{\odot}$\fi}
\def\msunyr{\ifmmode \mathrm{M}_{\odot} \hspace{1mm}{\rm yr}^{-1} \else $\mathrm{M}_{\odot}$ yr$^{-1}$\fi}
\def\zsun{\ifmmode Z_{\odot} \else Z$_{\odot}$\fi}
\def\lsun{\ifmmode L_{\odot} \else L$_{\odot}$\fi}
\def\mstar{\ifmmode \mathrm{M}_{\star} \else M$_{\star}$\fi}

\newcommand{\myemail}{eros.vanzella@oabo.inaf.it}

\slugcomment{}

\shorttitle{The first imaging of Lyman continuum emission from a distant star-forming galaxy}
\shortauthors{Vanzella et al.}

\begin{document}

\title{Hubble imaging of the ionizing radiation from a star-forming galaxy at $z=3.2$ with $f_{esc}>50$\%}

\author{E. Vanzella\altaffilmark{1,*}, S. de Barros \altaffilmark{1}, K. Vasei \altaffilmark{2}, A. Alavi \altaffilmark{2}, M. Giavalisco \altaffilmark{3},
B. Siana \altaffilmark{2}, A. Grazian \altaffilmark{4}, G. Hasinger \altaffilmark{5}, H. Suh \altaffilmark{5}, N. Cappelluti \altaffilmark{6}, F. Vito \altaffilmark{7},
R. Amorin \altaffilmark{4}, I. Balestra \altaffilmark{8,9}, M. Brusa \altaffilmark{10,1}, F. Calura \altaffilmark{1}, M. Castellano \altaffilmark{4}, 
A. Comastri \altaffilmark{1}, A. Fontana \altaffilmark{4}, R. Gilli \altaffilmark{1},
M. Mignoli \altaffilmark{1}, L. Pentericci \altaffilmark{4}, C. Vignali \altaffilmark{10} and G. Zamorani \altaffilmark{1} 
}

\altaffiltext{1}{INAF--Osservatorio Astronomico di Bologna, via Ranzani 1, 40127 Bologna, Italy}
\altaffiltext{2}{Departement of Physics and Astronomy, University of California, Riverside, CA 92507, USA}
\altaffiltext{3}{Astronomy Department, University of Massachusetts, Amherst, MA 01003, USA}
\altaffiltext{4}{INAF--Osservatorio Astronomico di Roma, via Frascati 33, 00040 Monteporzio, Italy}
\altaffiltext{5}{Institute for Astronomy, 2680 Woodlawn Drive, Honolulu, HI 96822, USA}
\altaffiltext{6}{Yale Center for Astronomy and Astrophysics, Physics Department, New Haven, CT 06520, USA}
\altaffiltext{7}{Department of Astronomy and Astrophysics, 525 Davey Laboratory, The Pennsylvania State University, University Park, PA 16802, USA}
\altaffiltext{8}{INAF - Osservatorio Astronomico di Trieste, via G. B. Tiepolo 11, I-34131, Trieste, Italy}
\altaffiltext{9}{University Observatory Munich, Scheinerstrasse 1, D-81679 M\"unchen, Germany}
\altaffiltext{10}{Dipartimento di Fisica e Astronomia, Università degli Studi di Bologna, Viale Berti Pichat 6/2, 40127 Bologna, Italy}

\altaffiltext{*}{\myemail}

\begin{abstract}
Star-forming galaxies are considered to be the leading candidate sources that dominate the cosmic reionization
at $z>7$, and the search for analogs at moderate redshift showing Lyman continuum (LyC)
leakage is currently a active line of research. 
We have observed a star-forming galaxy at $z=3.2$ with {\it Hubble}/WFC3 in the F336W filter, 
corresponding to the 730-890\AA~rest-frame, and detect LyC emission. This galaxy is very compact and also has 
large Oxygen ratio \oiiiv/\oiilam\ ($\gtrsim 10$).
No nuclear activity is  revealed from optical/near-infrared spectroscopy and deep multi-band photometry 
(including the 6Ms X-ray {\it Chandra}). 
The measured escape fraction of ionizing radiation spans the range 50-100\%, depending on the IGM attenuation.
The LyC emission is detected with $m_{F336W}=27.57\pm0.11$ (S/N=10) and it
is spatially unresolved, with effective radius $R_e<200$pc.
Predictions from photoionization and radiative transfer 
models are in line with the properties reported here, indicating that stellar winds and supernova explosions in a nucleated
star-forming region can blow cavities generating density-bounded conditions compatible with optically thin media. 
Irrespective to the nature of the ionizing radiation, spectral signatures of these sources over the entire electromagnetic 
spectrum are of central importance for their identification during the epoch of reionization, 
when the LyC is unobservable. Intriguingly, the {\it Spitzer}/IRAC  photometric signature of intense rest-frame optical
emissions (\oiiidoub\ + \hb) observed recently at $z\simeq7.5-8.5$ is similar to what is observed in this galaxy. 
Only the {\it James Webb Space Telescope} will measure optical line ratios at 
$z>7$ allowing a direct comparison with lower redshift LyC emitters, as reported here.
\end{abstract}

\keywords{galaxies: evolution ---  galaxies: high-redshift --- dark ages, reionization, first stars}

\section{Introduction}
\label{sec:intro}

Cosmic reionization is a major episode in the history of the Universe and the search for ionizing sources is one of the
main goals of modern observational cosmology (Robertson et al. 2010).
Star-forming galaxies and AGN have been proposed to be the dominant sources of ionizing radiation, possibly
acting at different cosmic epochs (Haardt \& Madau 2012). While the redshift evolution of the ultraviolet
luminosity function of star-forming galaxies is relatively well measured up to $z\simeq 7-8$, 
showing that the bulk of the ultraviolet luminosity density is dominated by the faint galaxy 
population \citep[$L<L^{\star}$, e.g,][]{bouwens+15}, the number density of high redshift and faint AGN is still
highly uncertain \cite[e.g.,][]{giallongo+15,georgakakis+15}.
In addition, understanding the mechanisms of reionization (and post-reionization $z<6$) hinges on understanding
how the escape fraction of ionizing radiation, \flyc, 
changes as a function of luminosity and redshift. 
Because of the IGM opacity, direct observation of ionizing radiation during reionization is
not feasible \citep{prochaska+10}. 
A strategy to make progress is to identify LyC sources at lower redshift, e.g. 3-3.5,
and study which of their properties can be used as predictors of a LyC leakage.
Recent advances have been made by looking at starburst galaxies in the
local universe \citep{borthakur+14,izotov+16}, about 10 Gyrs after reionization ended ($z\sim6$). 
However, it is more useful to identify LyC leakers at the highest redshifts possible ($z\sim3$,
one billion years after the end of reionization), because the properties of these galaxies
are better analogs to those that reionized the universe. 

Escaping LyC radiation from galaxies has been searched in the recent years 
\citep{vanzella+10b,siana+10,nestor+13,mostardi+13,mostardi+15}
and, to date, no convincing LyC emission has been reported at high redshift \citep{siana+15}. 
Moreover, nothing is known about the spatial distribution of the emerging LyC radiation. 
Here we report on a LyC emission arising from a distant galaxy ($z=3.212$) unambiguously confirmed with
\hst\ observations.

Throughout the paper, the
AB system ($AB=31.4 - 2.5log(f_\nu/nJy$) and cosmology 
$\Omega_{tot}$, $\Omega_M$, $\Omega_\Lambda = 1.0, 0.3, 0.7$ with
$H_0 = 70$kms$^{-1}$Mpc$^{-1}$ are used. 

%
\begin{figure}[htbp]
\centering
\includegraphics[width=8.5cm,trim=0cm 0cm 0cm 0cm,clip=true]{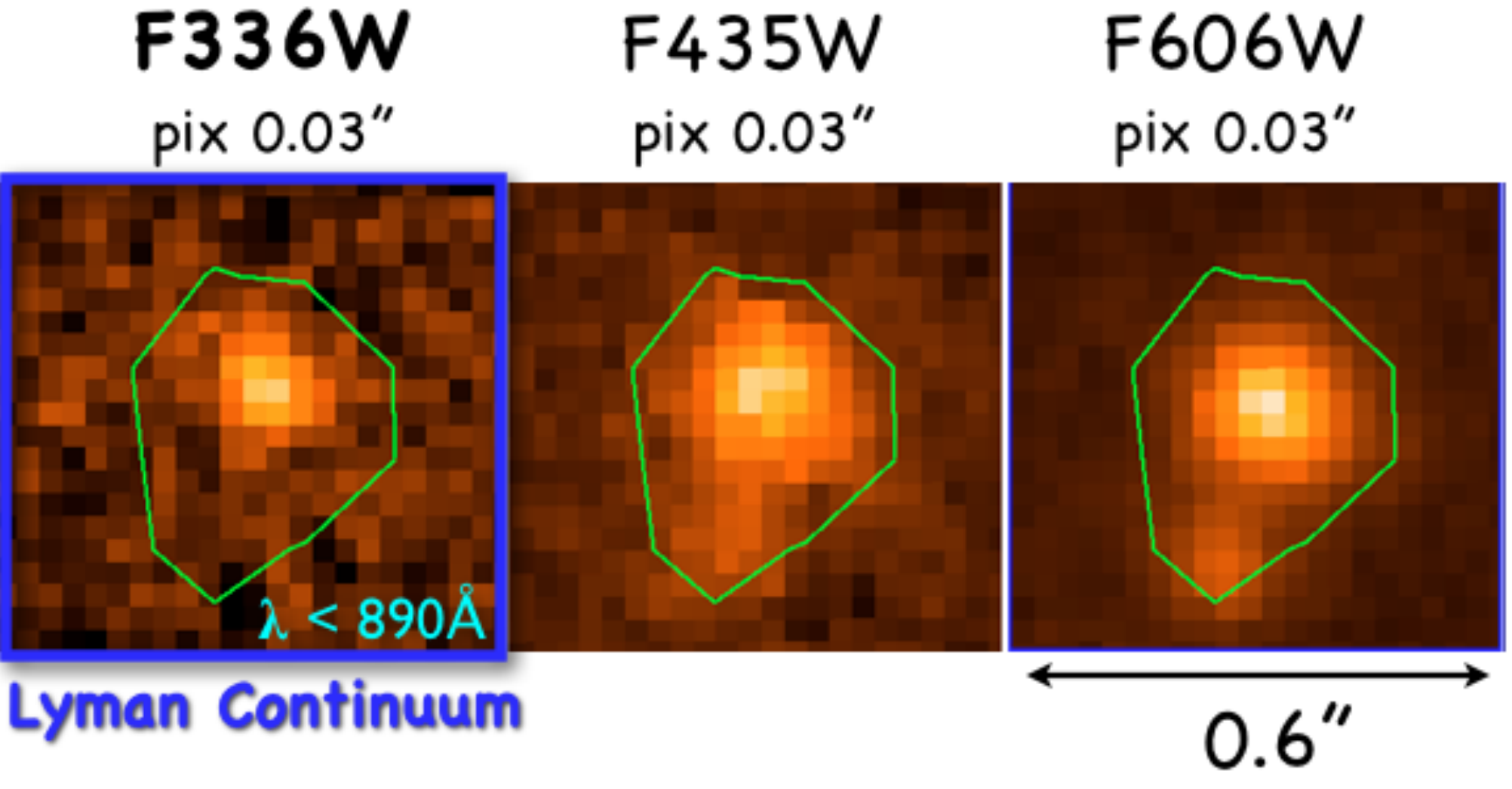}
\caption{From left to right, the F336W, F435W and F606W \hst/ACS thumbnails of \ionii. A LyC emission
(rest-frame $\lambda<890$\AA) arising from the brighter component (A) is evident in the F336W-band images. The green
iso-contour derived from the F435W band guides the eye in comparing the shape of the
source. 
Secondary emission (B), offset by 0.2\arcsec, is visible below the main source in the F435W and F606W images, but not in the F336W.}
\label{fig:fig1}
\end{figure} 

\section{The challenging detection of escaping LyC photons}

The difficulty of directly and unambiguously identifing the LyC radiation in distant galaxies are due to a combination of several effects: 
(1) superposition of foreground sources can produce false LyC detections \citep{vanzella+10b,vanzella+12}, and in this respect
{\it HST} observations are crucial \citep{siana+15}; (2) the intergalactic transmission in the ionizing 
domain is stochastic \citep{inoue+14}; (3) the geometrical distribution of the neutral gas in galaxies and the relatively short duty cycle 
of \flyc\ over cosmic time adds further stochasticity in the LyC visibility \citep{wise+14,cenkimm2015}; 
(4) the accessible luminosity range in the LyC achievable by current large telescopes limits the analysis to bright galaxies 
($L>0.5L^\star$) for which the average escaping ionizing radiation has now been shown to be intrinsically low 
\citep[$<5-10\%$,][]{grazian+16,vanzella+10c,mostardi+15,siana+15}. 
These aspects make the search for escaping ionizing radiation challenging and can explain the current low detection
rate at high redshift.
Nevertheless, the identification of examples with escaping LyC radiation lying in the tail of the large \flyc\ values, though rare 
\citep{vanzella+10c}, represents the only empirical way we have to increase our physical insight into the mechanisms that allow 
ionizing photons to escape and thus providing unique reference for studies at $z>6$.

\subsection{The selected candidate LyC emitter}
A LyC candidate at $z=3.212$ in the GOODS-Southern field, named \ionii\ (GDS-ID~033203.24-274518.8), 
was first identified by \cite{vanzella+15}. 
The compact star-forming region, showing strong \oiiidoub\ emission lines (with a rest-frame equivalent width of 1500\AA) and 
large Oxygen ratio \oiiiv/\oiilam\ ($\geq10$) makes \ionii\ the highest redshift ``Green Pea'' galaxy currently known and, accordingly
to the photoionization models \citep{nakajimaouchi2014,jaskotoey2013}, an ideal candidate LyC emitter. 
A plausible spectroscopic LyC detection was subsequently
discussed in \cite{debarros+16}. However, the presence of a close companion not resolved with ground-based spectroscopy
and imaging ($0.2''$, see Fig.~\ref{fig:fig1}) casted some doubts about the association with \ionii\ of the observed flux and 
thus on the reliability of the LyC leakage.

\section{\hst\ observations}

To confirm the LyC emission, we obtained \hst\ WFC3/UVIS image in the F336W filter, corresponding
to rest-frame $730-890$ \AA. Seventeen 2800s (1-orbit) dithered exposures were taken for a total integration time of 47.6 ks. The charge
transfer efficiency (CTE) of the WFC3/UVIS CCDs has degraded significantly due to radiation damage and is particularly problematic
in images with low background. To mitigate the effects of poor CTE, we increased the background to $\sim12 e^-$ pix$^{-1}$ with a
post-flash LED \citep{birettabaggett13}. Furthermore, we placed the target near the read-out edge of the CCD so that electrons
need only transfer across $\sim300$ of the 2048 pixels. Finally, we applied a pixel-based CTE
correction\footnote{http://www.stsci.edu/hst/wfc3/tools/cte\_tools}.
As the dark current is nearly half of the total background, proper dark subtraction is critical. Unfortunately, the STScI darks are only
a single value and do not capture the gradient and blotchy pattern in the dark current \citep{teplitz+13}. Also, because the darks
are not CTE corrected, more than half of hot and warm pixels are not properly masked \citep{rafelski+15}. To mitigate these issues
we adopted the dark processing method explained in detail in \cite{rafelski+15}. First, we CTE-correct all raw dark images in the 
anneal cycle of our visits and remove the cosmic rays. We then find and mask the hot pixels that appear throughout the anneal cycle 
and make a mean super dark from all darks in the anneal cycle. We then dark subtract the science images while masking the hot pixels 
that were found to exist at the time of the observation. 

We processed all of our CTE-corrected raw data using STSDAS task CALWF3, including subtraction of our new super darks. These 
calibrated images were then combined using Astrodrizzle \citep{gonzaga+12}, which performs background subtraction, cosmic 
ray rejection and geometric distortion correction. The final combined F336W image has a pixel scale of 0.03\arcsec~and is 
astrometrically aligned with the 3D-HST F606W image \citep{skelton+14} with a precision of 48 mas. As a by-product, 
Astrodrizzle produces an inverse variance image used to derive the uncertainties in the photometry.
We performed aperture photometry using a growth-curve methodology. The estimated uncertainty was corrected for the correlated 
noise introduced in the drizzling process \citep{casertano+00}. Finally, we corrected our photometry for 
Galactic extinction \citep[0.041 magnitude;][]{schlaflyfinkbeiner11}. 

\begin{figure}[htbp]
\centering
\includegraphics[width=8.5cm,trim=0cm 0cm 0cm 0cm,clip=true]{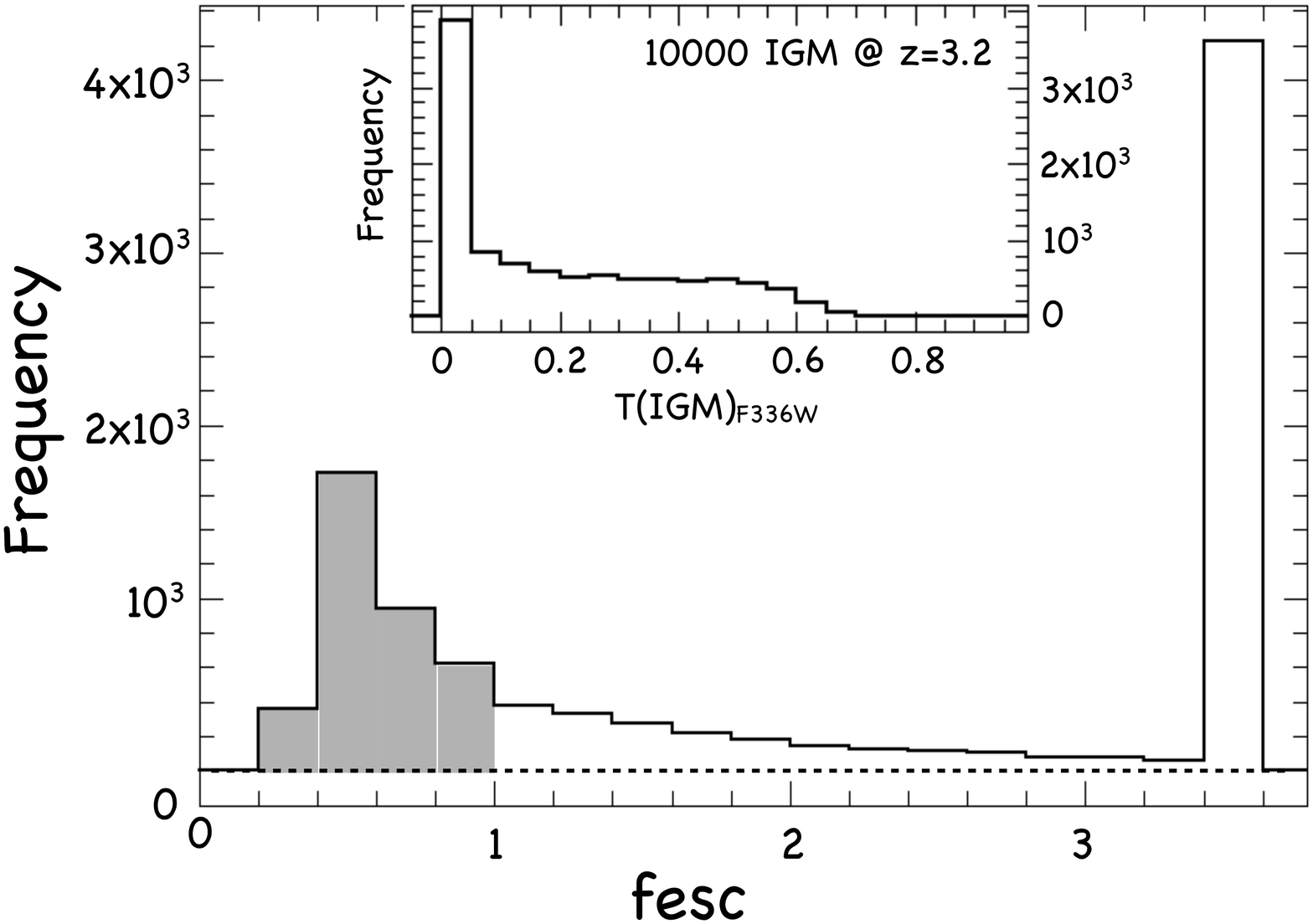}
\caption{The distribution of  \flyc\ is shown in the main panel adopting
  an intrinsic luminosity density ratio (L1500/L800)$_{INT}$=5 (see Eq.~1). Values of \flyc~$ > 3.5$ have fixed
  to 3.5 to better display the entire distribution. \flyc~$ < 1$ corresponds to 37\% of the 10000 realizations (gray histogram). 
  The inset shows the IGM transmission at $z=3.2$ (from \citet{inoue+14}).
  The F336W flux requires both a relatively high  \flyc\ ($>40\%$) or small ratio (L1500/L800)$_{INT}<10$.
  }
\label{fig:fig2}
\end{figure}

\section{Results}
\label{sec:result}

The HST imaging clearly solved the problem of the close neighbour and the LyC emission has been
unambiguously confirmed at $S/N=10$ with magnitude $m(\mathrm{F336W})=27.57 \pm0.11$
(see Fig.~\ref{fig:fig1}). It also provides the spatial mapping of ionizing
radiation from a galaxy for the first time.
From deep multi-wavelength observations, including {\it Chandra} 6Ms image, \hst\ optical \citep[GOODS,][]{giavalisco+04b} and 
near-infrared \citep[CANDELS,][]{grogin+11}, {\it Spitzer} (3.6, 4.5, 5.8, 8.0, 24\micron) and wide spectroscopic 
coverage from $U$ to the $K$ band (VLT and Keck), we found \ionii\ to be a low-mass 
($\leq10^9\msun$), low-metallicity ($\sim1/6\zsun$ ) star-forming galaxy \citep{debarros+16}.

We derived the \flyc\  quantity with usual formulation \citep[e.g.,][]{vanzella+12}:
\begin{equation}
\flyc = \frac{\left( L1500/L800\right)_{INT}}{\left( f1500/f800 \right)_{OBS}} \times \frac{1}{T(IGM)_{F336W}}\times 10^{-0.4\times A1500}
\end{equation}
where f1500/f800 is the observed flux density ratio ($=14.60$) and the dust attenuation A1500 has been derived by \citet{debarros+16} 
(E(B-V)~$<0.04$, A1500~$\simeq0.4$).  Fig.~\ref{fig:fig2} shows the distribution of \flyc\ calculated adopting the intrinsic ratio
(L1500/L800)$_{INT}=5$ \citep{siana+07}  and convolving with 10000 IGM transmission at $z=3.2$ (Eq.~1).
A \flyc~$<1$ is still possible if (L1500/L800)$_{INT}<15$ with a high T(IGM)$_{F336W}$ ($\simeq0.7$). 

\subsection{The ionizing photon production rate}
The measured ionizing photon production rate corresponding to the observed $m(\mathrm{F336W})=27.57$ is 
$N_\mathrm{phot}(800$\AA$)=1.7\times10^{53}\mathrm{s}^{-1}$. The statistical error related to the measured magnitude is negligible if compared to the 
stochasticity of the IGM attenuation affecting a single line of sight, as shown by the distribution of F336W--IGM convolved transmissions
derived from the IGM prescription of \citet{inoue+14} ($T(IGM)_{F336W}$, Fig.~\ref{fig:fig2}). 
Assuming the maximum $T(IGM)_{F336W}$ allowed at $z=3.2$
($\simeq 70$\%), we have an intrinsic $m(\mathrm{F336W})=27.18$ that corresponds
to $N_\mathrm{phot}(800$\AA$)=2.5\times10^{53}\mathrm{s}^{-1}$. 
We can set a conservative upper limit to $N_\mathrm{phot}$ by assuming that the intrinsic F336W magnitude (ionizing emission)
cannot be brighter than the observed magnitude at 1500\AA\ rest-frame ($\sim24.66$), i.e. the intrinsic luminosity
density ratio is $L1500/L800>1$ (i.e., no extreme stellar populations are present). We obtain 
$N_\mathrm{phot}(800$\AA$)=2.5\times10^{54}\mathrm{s}^{-1}$. Therefore the intrinsic ionizing photon production rate 
is $2.5\times10^{53}\mathrm{s}^{-1}\leq N_\mathrm{phot}(800$\AA$)\leq 2.5\times10^{54}\mathrm{s}^{-1}$.

\subsection{LyC morphology}
\begin{figure}[htbp]
\centering
\includegraphics[width=8.5cm,trim=0cm 0cm 0cm 0.0cm,clip=true]{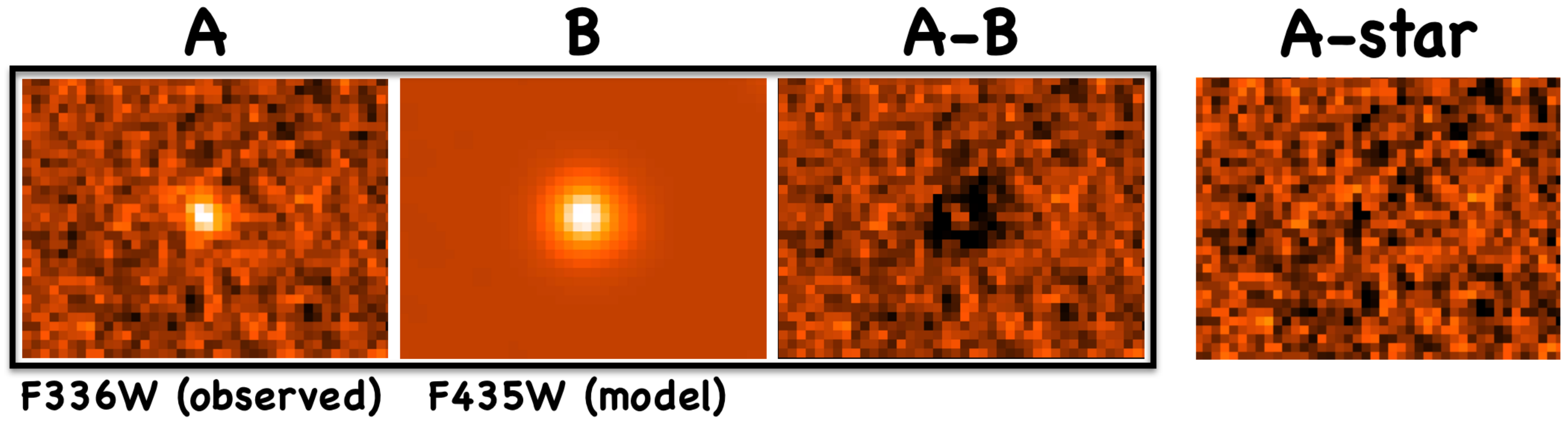}
\caption{The difference between the observed F336W image (A) {\sc Galfit} best-fit model calculated in the F435W image (B) is shown in the right, (A-B). The model A is scaled to the flux peak of the F336W source (B). This comparison suggests the LyC emitting region is more compact than the resolved emission measured in the F435W band.  In the right panel the
residuals after subtracting a star rescaled to the source flux is shown.
The size of the thumbnails is  $1.2''\times1.0''$.}
\label{fig:fig4}
\end{figure}

\begin{figure}[htbp]
\centering
\includegraphics[width=8.5cm,trim=0cm 0cm 0cm 0cm,clip=true]{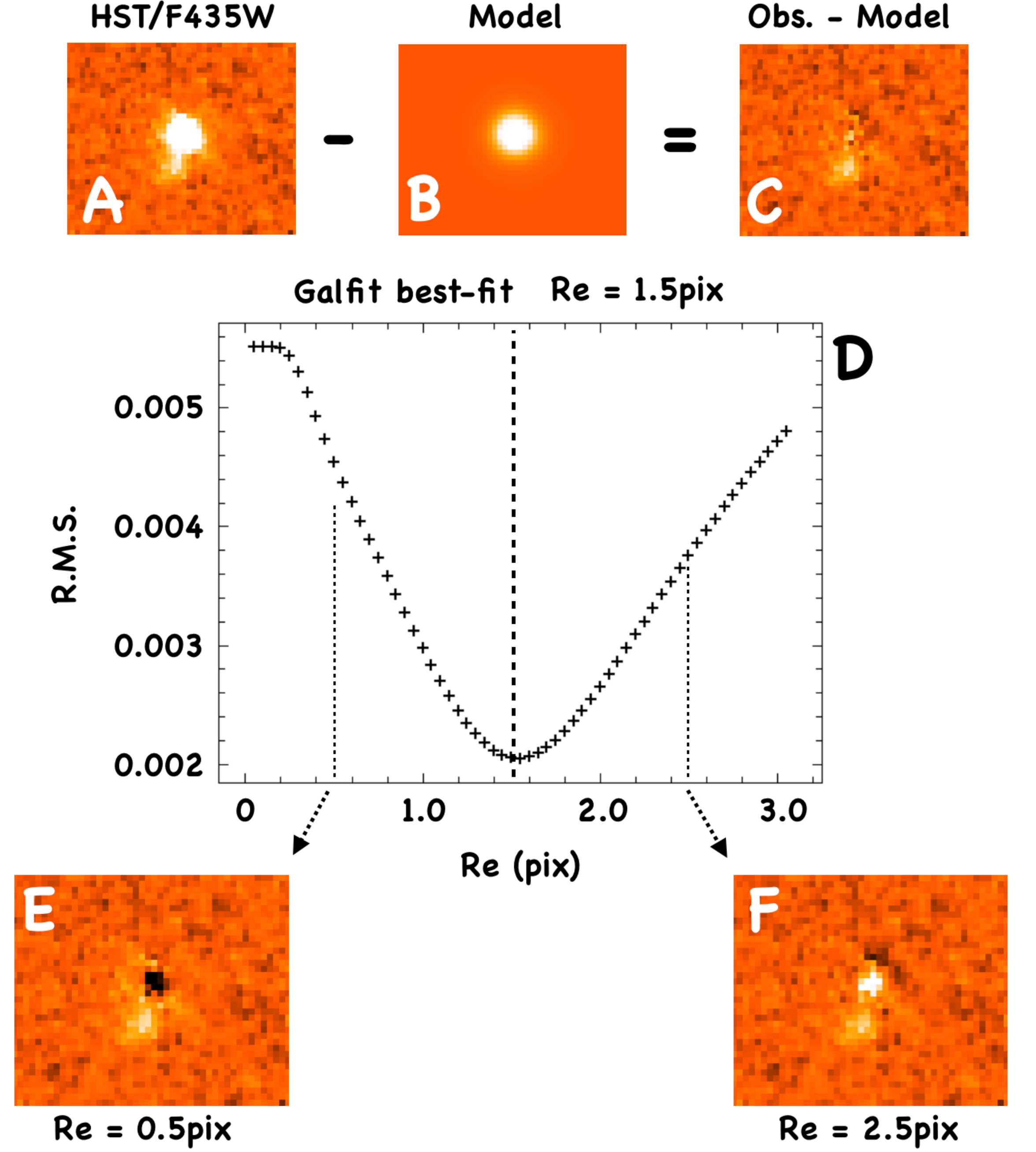}
\caption{{\sc Galfit} modeling of the \ionii\ galaxy in the F435W band (1000\AA\ rest-frame). In
panel C the residuals of observed (A) - model (B) are shown for the best-fit case ($R_e=1.5px \simeq 340$pc). 
Bottom panels show an example of two ``bad'' solutions 
(observed-model), in which positive/negative residuals  are evident (panels E/F). 
Middle panel (D) shows the behaviour of the r.m.s. calculated at the `A' position by running {\sc Galfit} in a grid of effective radii 
(as described in \citet{vanzella+15}). A good convergence is reached at $R_e=1.5\pm0.1$px. 
This illustrates that the source is spatially resolved at 1000\AA\ rest-frame. The size of the thumbnails is $1.2'' \times1.0''$. }
\label{fig:fig3}
\end{figure}

The LyC leakage is co-spatial with the component $A$ and nothing is detected from component $B$,
where $A$ and $B$ refer to the brighter and the fainter blobs, respectively (see Fig.~\ref{fig:fig1},
following the \citet{vanzella+15} nomenclature). 
\ionii\ is a compact but well resolved galaxy in the F435W, F606W, F775W and F850LP ACS images, 
corresponding to the rest-frame far-UV from 1030 to 2020 \AA. The galaxy, however, does not appear 
to be resolved in the WFC3 F336W image. We have a well-exposed nearby bright star and inserted 
in a blank area of the image at random position to simulate a number ($N=50$) of realizations of 
the PSF, after rescaling it to the same flux as \ionii. We have then registered the star at the same 
position of the light centroid of \ionii\ and subtracted it, in each case obtaining residuals consistent 
with the sky background (an example is shown in Fig.~\ref{fig:fig4}). 
We have also compared the morphology of \ionii\ in the F336W and F435W 
bands to test the possibility that the lower SNR in the former band is the reason of the apparent 
unresolved morphology, adopting the following procedure.
Basic PSF-corrected morphological information from both components were
extracted by \cite{vanzella+15} using {\sc Galfit} \citep{peng+10}. Given their regular (symmetric) morphology, the {\sc Galfit} modeling
with a Gaussian light profile well reproduces both regions in all the ACS images and shows they are spatially resolved 
with effective radii of a few hundred parsecs.
In particular, we show in Fig.~\ref{fig:fig3} the result obtained at 1000\AA\ rest-frame  
(F435W band) adopting a Gaussian shape. The best solution produces an effective radius of 
$R_e=340\pm 25$pc, suggesting the stellar radiation emerges from a compact and resolved 
region; the error has been calculated
by running {\sc galfit} on simulated images obtained by inserting the B-band model in random and free regions of the F435W-band image. 
{\sc Galfit} fitting on the LyC image does not produce resolved solutions.
In order to characterize the LyC morphology we perform a simple test by subtracting the best-fit B435-band model 
from the F336W image.  
After normalizing the two images at the same flux peak and checking the residuals also
over a grid of dimming factors, it turns out that the LyC emission arises from a region
smaller than what has been inferred from 
the F435W band (see Fig.~\ref{fig:fig4}),
in particular the spatially unresolved detection in F336W corresponds to an effective radius $< 200$pc.
The negative residual flux seen in Fig.~\ref{fig:fig4} extends at least 5 pixels
radially from the center, and can't be explained by the small variations in the PSFs at the two wavelengths.
The observed compactness of the LyC emission is probably not surprising 
if we compare our \hst\ resolution to the size of the super-star clusters observed
in local starbursts, in which the O-type stars are spatially segregated toward the center within tens of parsecs
\citep[e.g.,][]{annibali+15,james+16}.

Another possible interpretation is that the LyC emission of \ionii\ could originate from an AGN. This is unlikely and is
discussed in Sect.~\ref{sec:agn}.

\begin{figure*}[htbp]
\centering
\includegraphics[width=18cm,trim=0cm 0cm 0cm 1cm,clip=true]{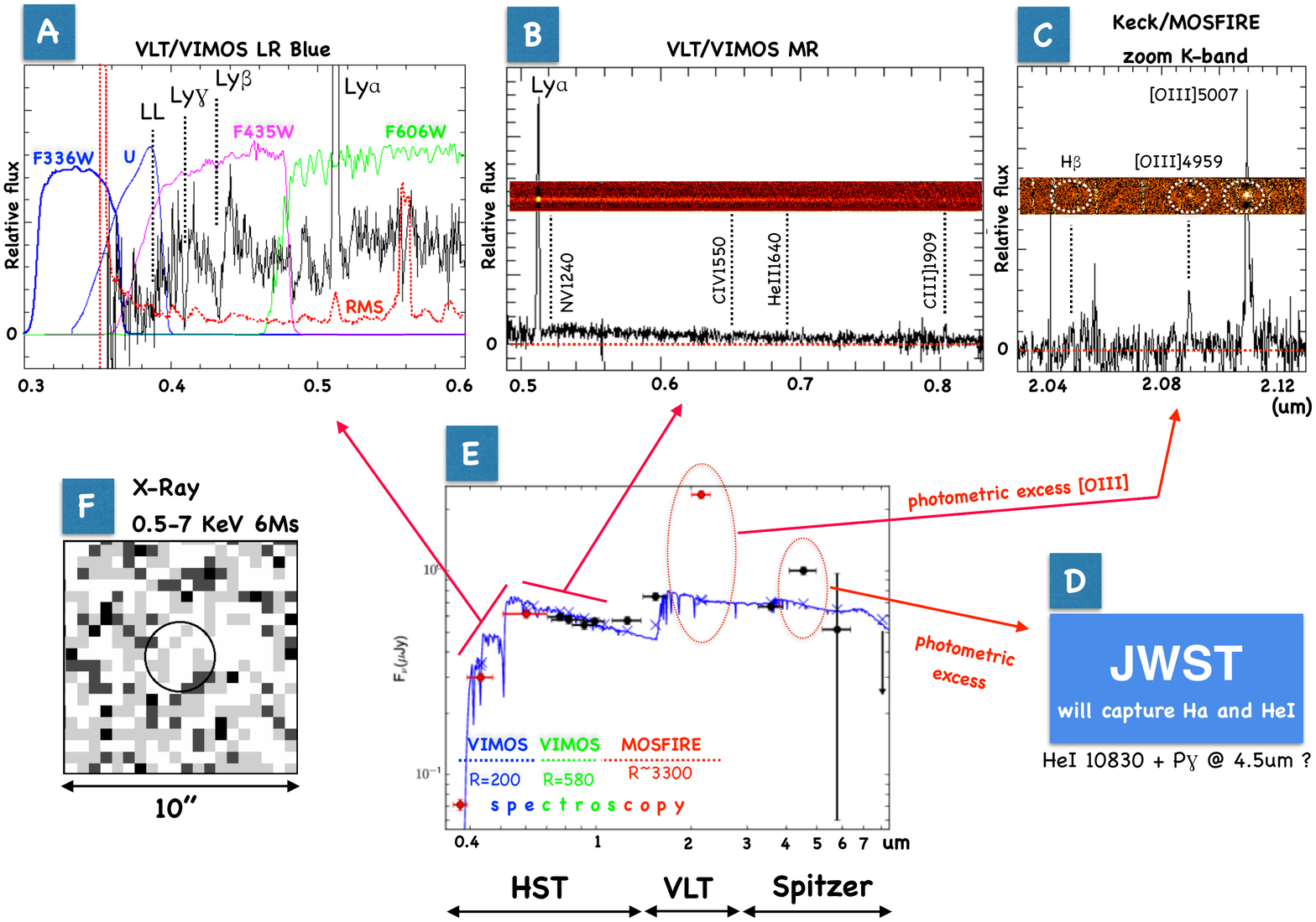}
\caption{A panoramic view of the available multi-frequency data for \ionii, from the 0.35 to $2.4\mu m$ wavelength range (VLT/VIMOS and Keck/MOSFIRE). {\it Panels A, B and C}: spectral coverage from U to the K-band with VLT/VIMOS and Keck/MOSFIRE. The insets of panel B and C show the two-dimensional spectra.The absence of high ionization emission lines (\nv, \civ, \heii) is evident, as well as the clear detection of \lya\ and \ciii\ lines. The \oiiidoub\ lines from Keck/MOSFIRE spectrum are also shown \citep[from][]{debarros+16}. In panel E a schematic view of the spectral SED fitting indicating the available observations (VLT, Keck, HST, Spitzer) is shown. JWST can cover the part at 2.4-5\micron, including lines as \ha\ and \hei, not accessible with current instruments (panel D).
Panel F shows the 6Ms X-ray 0.5-7 keV Chandra cutout (10$\arcsec\times$10$\arcsec$) around \ionii\ obtained from the public Chandra observation
(the circle of diameter  $3\arcsec$ marks the position of \ionii). The source is off-axis  thus the PSF of the source would cover the entire circle symbol.} 
\label{fig:fig5}
\end{figure*}

\section{Discussion}
\label{sec:discuss}

Independent from the nature of ionizing photons, the LyC detection implies a low column density of neutral gas along the
line of sight, lower than $10^{17.2}$ cm$^{-2}$ (corresponding to $\tau(\mathrm{LyC})<1$). 
\cite{debarros+16} has shown for the first time empirical evidences linking high \flyc\ with compactness of the star-forming region, the large \oiii/\oii\ 
line ratio, the weak ultraviolet interstellar absorption lines, and the emerging \lya\ emission close to the systemic redshift, as predicted by 
photoionization and radiative transfer models when the medium is considered optically thin 
\citep{verhamme+15,jaskotoey2013,borthakur+14,nakajimaouchi2014,izotov+16}. The \hst\ observations presented here unambiguously confirm those predictions.

\subsection{Nature of the source of the LyC photons}
\label{sec:agn}

In the following, we summarize a few key elements \citep{debarros+16} and also add new empirical and theoretical evidence
that favors the stellar origin of the ultraviolet light.
\begin{enumerate}

\item No high-ionization emission lines like \nv, \civ, \heii\ have been detected at more than 3$\sigma$ level from the VLT/VIMOS MR spectrum 
(Fig.~\ref{fig:fig5}), while the \ciii\ line is clearly detected. From $\civ/\ciii<0.15$ and $\ciii/\heii>4.0$ line ratios and following \cite{feltre+16}, this source is 
classified as a star-forming galaxy, lying in the same region occupied by low-metallicity galaxies of \cite{stark+14b}.

\item \ionii\ is spatially resolved in all of the ACS bands ($340\pm25$pc) implying the stellar emission is detected  and dominating the observed
range 1000\AA-2000\AA\ rest-frame.

\item  The narrow width of the \oiiiv\ emission line ($\sigma=65$km s$^{-1}$) is compatible with lower redshift star-forming galaxies 
\citep{maseda+14}, while the AGN population typically show higher velocity dispersions up to several hundreds or thousand km s$^{-1}$.

\item \ionii\ is not detected in the X-ray in the 4 Ms CDFS (\citet{cappelluti+15})
and no evidence of emission is inferred from the recent Chandra pointings publicly available on the Chandra website\footnote{http://cxc.harvard.edu/}, 
which increased the exposure time to $\sim6$Ms (see Fig.~\ref{fig:fig5}). This places an X-ray luminosity limit of 
$L_X \lesssim 3\times10^{42}~erg s^{-1}$ at $1\sigma$ limit. 
If \ionii\ were a type I AGN, and assuming that the $L_X-L_{\oiii}$ relation observed locally is valid at $z=3$, the measured \oiiiv\ luminosity \citep{debarros+16} would 
imply a detection with $S/N>100$ at the \ionii\ position in the 6Ms X-ray image, corresponding to $L_X \simeq 10^{45}~erg s^{-1}$ \citep{panessa+06,ueda+15}. 
The measured \oiiiv\ line luminosity ($2\times10^{43}~erg s^{-1}$) and the non detection in the 6Ms X-ray would imply a high obscuration
(if it were an AGN), formally with an equivalent column density $N_H>10^{25}cm^{-2}$ (assuming solar metallicity). 
The estimated stellar mass from the LyC 
source is $\sim10^8~\msun$ (see Section~\ref{sec:sb99}) and, following \cite{maseda+14}, the dynamical mass from the \oiiiv\ line width 
($\sigma=65~km s^{-1}$) is $\sim10^9\msun$.
A lower limit on the BH mass of $10^8~\msun$ can be estimated by converting the \oiiiv\ line luminosity into an AGN bolometric
luminosity \citep{panessa+06,lusso+12}  and then assuming that the AGN is radiating at the Eddington limit, 
that imply a very unusual and probably unrealistic ratio of $M_{BH}$/Total mass $\geq0.1$ for this object.

\item Adopting the expected $L_X$  (from  \oiiiv\ reported above), we expect a clear detection at 6\micron\ rest-frame (i.e., at 24\micron\ observed with {\it Spitzer}/MIPS) by 
assuming valid the relation between $L_X$ and $L(6\micron)$ \citep{stern+15}. In particular even adopting a $L_X>10^{44}$erg s$^{-1}$ we expect 
$L$(6\micron)$>10^{44}$erg s$^{-1}$ that corresponds to mag$(24$\micron)$<20$(AB) and therefore a detection in the MIPS 24\micron\ image 
at $S/N>10$. 
At the position of \ionii\ no signal is detected in the MIPS $24\mu m$ band down to AB$\simeq22.3$ \citep{debarros+16}. 
\end{enumerate}

\noindent It is worth noting a photometric excess ($S/N>5$) in the second IRAC channel at $4.5\micron$ (Fig.~\ref{fig:fig5}),
that we ascribed to possible \hei+\pg\ lines with rest-frame equivalent width of $\sim 1100$\AA. While these lines can only be observed with 
{\it JWST}, such a strong \hei\ emission up to 1000\AA\ equivalent width has been observed in local compact \hii\ regions
by \cite{izotov+14} and in a local LyC emitter candidate \citep{verhamme+15}.

Therefore we conclude that the signal observed in the \hst/F336W band is very likely due to stellar emission;
the only (unlikely) alternative would be the presence of a heavily obscured AGN,
located at the center of the star-forming region and hidden by gas and dust at all wavelengths but visible in
the ionizing continuum, a possibility we deem very contrived. 
An intrinsically faint AGN could co-exist with a dominating star formation activity that controls the transparency of the medium.

\subsection{Constraining the burst of star formation}
\label{sec:sb99}

The measured ionizing photon production rate and the comparison with Starburst99 models \citep{leitherer+99} adopting
instantaneous burst, $Z=0.004$ and Salpeter IMF, provide a stellar mass involved in the starburst event in the range
$5\times10^6-5\times10^7\msun$ (depending on the IGM transmission), with the young stellar component 
($<10$Myr) dominating the ionizing radiation, in which the number of O-type stars ranges between $2\times10^4 - 2\times10^5$. 
The ionizing photon production rate is similar to the one derived at lower redshift for a recently discovered LyC emitter \citep{izotov+16}.

Such a young starburst event can generate substantial ionizing radiation and produce ionized cavities, and recent supernovae and stellar winds may 
have carved holes in the ISM that favour LyC photon escape into the intergalactic medium. It is worth stressing that the observed large \oiii/\oii\ line 
ratio ($>10$) is expected in this scenario \citep{jaskotoey2013,nakajimaouchi2014}.
Similar large \oiii/\oii\ ratios have been observed in local starbursts \citep{james+16} in which the younger stellar component containing O-type stars is 
identified in the core of the starburst (similarly to what is inferred here) and can photoionize regions out to hundreds of parsecs, as observed
also in a local starburst by \citet{annibali+15}.
It is worth noting that the differential depression of nebular emission among Balmer and metal lines when a substantial
leakage of ionizing radiation is present could strongly affect the usual diagnostic diagrams that separate
star-forming vs. AGN emission \citep[e.g., BPT,][]{baldwin+81}; 
in particular, a LyC emitter would move toward the AGN cloud if the Balmer lines are attenuated first.

\section{Conclusions}
\label{sec:conclusion}    

While the LyC emitter reported here is rare among the sources with similar luminosity \citep{vanzella+10c,grazian+16}, the non-ionizing multi-frequency
properties observed in our galaxy and the confirmed LyC emission provide valuable prospects for the characterization of similar or fainter sources in
higher redshift domains. In particular, the confirmed $z>7.5$ galaxies \citep{finkelstein+13,oesch+15,zitrin+15} show particularly strong Oxygen and Balmer structure (\oiiidoub\ + \hb), at the same level reported here (equivalent widths larger than 800-1000\AA\ rest-frame). 
It is premature to address if those sources have effectively an $\flyc>0$, as optical line ratios are needed
to perform a direct comparison and this is postponed until {\it JWST} launch. Our result is currently a unique high-redshift reference,
both in terms of large Oxygen line ratio and large line equivalent width, and needs to be extended to statistically significant samples,
especially investigating the faint luminosity domain. 
It also outlines the feasibility of the identification of ionizing sources during the reionization epoch.

\acknowledgments

We thank F. Annibali and D. Schaerer for useful discussions.
Part of this work has been funded through the INAF grants (PRIN INAF 2012).
MB acknowledges support from the FP7 Grant ``eEASy'' (CIG 321913).


\begin{thebibliography}{47}
\expandafter\ifx\csname natexlab\endcsname\relax\def\natexlab#1{#1}\fi

\bibitem[{{Annibali} {et~al.}(2015){Annibali}, {Tosi}, {Pasquali}, {Aloisi},
  {Mignoli}, \& {Romano}}]{annibali+15}
{Annibali}, F., {Tosi}, M., {Pasquali}, A., {et~al.} 2015, ArXiv e-prints,
  arXiv:1505.05545

\bibitem[{{Baldwin} {et~al.}(1981){Baldwin}, {Phillips}, \&
  {Terlevich}}]{baldwin+81}
{Baldwin}, J.~A., {Phillips}, M.~M., \& {Terlevich}, R. 1981, \pasp, 93, 5, 5

\bibitem[Biretta \& Baggett (2013)]{birettabaggett13}
J. Biretta and S. Baggett, 2013, STScI, Instrument Science Report WFC3 2013-12, 
{\it http://www.stsci.edu/hst/wfc3/documents/ISRs}

\bibitem[{{Borthakur} {et~al.}(2014){Borthakur}, {Heckman}, {Leitherer}, \&
  {Overzier}}]{borthakur+14}
{Borthakur}, S., {Heckman}, T.~M., {Leitherer}, C., \& {Overzier}, R.~A. 2014,
  Science, 346, 216, 216

\bibitem[{{Bouwens} {et~al.}(2015){Bouwens}, {Illingworth}, {Oesch}, {Caruana},
  {Holwerda}, {Smit}, \& {Wilkins}}]{bouwens+15}
{Bouwens}, R.~J., {Illingworth}, G.~D., {Oesch}, P.~A., {et~al.} 2015, \apj,
  811, 140, 140

\bibitem[{{Cappelluti} {et~al.}(2015){Cappelluti}, {Comastri}, {Fontana}, {Zamorani},
{Amorin}, {Castellano}, {Merlin}, {Santini}, {Elbaz}, {Schreiber}, {Shu}, {Wang}, {Dunplo},
 {Bourne}, {Bruce}, {Buitrago}, {Micha{\l}owski}, {Derriere}, {Ferguson}, {Faber},
 {Vito}}]{cappelluti+15}
 {Cappelluti}, N.,  {Comastri}, A., {Fontana}, A., {Zamorani}, G.,      {Amorin}, R.,
 {Castellano}, M.,{et-al.}, 2015, \apj, submitted

\bibitem[{{Casertano} {et~al.}(2000){Casertano}, {de Mello}, {Dickinson},
  {Ferguson}, {Fruchter}, {Gonzalez-Lopezlira}, {Heyer}, {Hook}, {Levay},
  {Lucas}, {Mack}, {Makidon}, {Mutchler}, {Smith}, {Stiavelli}, {Wiggs}, \&
  {Williams}}]{casertano+00}
{Casertano}, S., {de Mello}, D., {Dickinson}, M., {et~al.} 2000, \aj, 120,
  2747, 2747

\bibitem[{{Cen} \& {Kimm}(2015)}]{cenkimm2015}
{Cen}, R., \& {Kimm}, T. 2015, \apjl, 801, L25, L25

\bibitem[{{de Barros} {et~al.}(2016){de Barros}, {Vanzella}, {Amor{\'{\i}}n},
  {Castellano}, {Siana}, {Grazian}, {Suh}, {Balestra}, {Vignali}, {Verhamme},
  {Zamorani}, {Mignoli}, {Hasinger}, {Comastri}, {Pentericci},
  {P{\'e}rez-Montero}, {Fontana}, {Giavalisco}, \& {Gilli}}]{debarros+16}
{de Barros}, S., {Vanzella}, E., {Amor{\'{\i}}n}, R., {et~al.} 2016, \aap, 585,
  A51, A51

\bibitem[{{Feltre} {et~al.}(2016){Feltre}, {Charlot}, \& {Gutkin}}]{feltre+16}
{Feltre}, A., {Charlot}, S., \& {Gutkin}, J. 2016, \mnras, 456, 3354, 3354

\bibitem[{{Finkelstein} {et~al.}(2013){Finkelstein}, {Papovich}, {Dickinson},
  {Song}, {Tilvi}, {Koekemoer}, {Finkelstein}, {Mobasher}, {Ferguson},
  {Giavalisco}, {Reddy}, {Ashby}, {Dekel}, {Fazio}, {Fontana}, {Grogin},
  {Huang}, {Kocevski}, {Rafelski}, {Weiner}, \& {Willner}}]{finkelstein+13}
{Finkelstein}, S.~L., {Papovich}, C., {Dickinson}, M., {et~al.} 2013, \nat,
  502, 524, 524


\bibitem[{{Giallongo} {et~al.}(2015){Giallongo}, {Grazian}, {Fiore}, {Fontana},
  {Pentericci}, {Vanzella}, {Dickinson}, {Kocevski}, {Castellano}, {Cristiani},
  {Ferguson}, {Finkelstein}, {Grogin}, {Hathi}, {Koekemoer}, {Newman}, \&
  {Salvato}}]{giallongo+15}
{Giallongo}, E., {Grazian}, A., {Fiore}, F., {et~al.} 2015, \aap, 578, A83, A83

\bibitem[{{Georgakakis} {et~al.}(2015){Aird}, {Buchner}, {Salvato}, {Menzel},
  {Brandt}, {McGreer}, {Dwelly}, {Mountrichas}, {Koki}, {Georgantopoulos},
  {Hsu}, {Merloni}, {Liu}, {Nandra}, \& {Ross},  \&}]{georgakakis+15}
{Georgakakis}, A., {Aird}, J., {Buchner}, J., {et~al.} 2015, MNRAS, 453, 1946, 1946

\bibitem[{{Giavalisco} {et~al.}(2004){Giavalisco}, {Ferguson}, {Koekemoer},
  {Dickinson}, {Alexander}, {Bauer}, {Bergeron}, {Biagetti}, {Brandt},
  {Casertano}, {Cesarsky}, {Chatzichristou}, {Conselice}, {Cristiani}, {Da
  Costa}, {Dahlen}, {de Mello}, {Eisenhardt}, {Erben}, {Fall}, {Fassnacht},
  {Fosbury}, {Fruchter}, {Gardner}, {Grogin}, {Hook}, {Hornschemeier}, {Idzi},
  {Jogee}, {Kretchmer}, {Laidler}, {Lee}, {Livio}, {Lucas}, {Madau},
  {Mobasher}, {Moustakas}, {Nonino}, {Padovani}, {Papovich}, {Park},
  {Ravindranath}, {Renzini}, {Richardson}, {Riess}, {Rosati}, {Schirmer},
  {Schreier}, {Somerville}, {Spinrad}, {Stern}, {Stiavelli}, {Strolger},
  {Urry}, {Vandame}, {Williams}, \& {Wolf}}]{giavalisco+04b}
{Giavalisco}, M., {Ferguson}, H.~C., {Koekemoer}, A.~M., {et~al.} 2004, \apjl,
  600, L93, L93

\bibitem[{{Gonzaga} \& {et al.}(2012)}]{gonzaga+12}
{Gonzaga}, S., \& {et al.} 2012,

\bibitem[{{Grazian} {et~al.}(2016){Grazian}, {Giallongo}, {Gerbasi}, {Fiore},
  {Fontana}, {Le F{\`e}vre}, {Pentericci}, {Vanzella}, {Zamorani}, {Cassata},
  {Garilli}, {Le Brun}, {Maccagni}, {Tasca}, {Thomas}, {Zucca},
  {Amor{\'{\i}}n}, {Bardelli}, {Cassar{\`a}}, {Castellano}, {Cimatti},
  {Cucciati}, {Durkalec}, {Giavalisco}, {Hathi}, {Ilbert}, {Lemaux}, {Paltani},
  {Ribeiro}, {Schaerer}, {Scodeggio}, {Sommariva}, {Talia}, {Tresse},
  {Vergani}, {Bonchi}, {Boutsia}, {Capak}, {Charlot}, {Contini}, {de la Torre},
  {Dunlop}, {Fotopoulou}, {Guaita}, {Koekemoer}, {L{\'o}pez-Sanjuan},
  {Mellier}, {Merlin}, {Paris}, {Pforr}, {Pilo}, {Santini}, {Scoville},
  {Taniguchi}, \& {Wang}}]{grazian+16}
{Grazian}, A., {Giallongo}, E., {Gerbasi}, R., {et~al.} 2016, \aap, 585, A48,
  A48

\bibitem[{{Grogin} {et~al.}(2011){Grogin}, {Kocevski}, {Faber}, {Ferguson},
  {Koekemoer}, {Riess}, {Acquaviva}, {Alexander}, {Almaini}, {Ashby}, {Barden},
  {Bell}, {Bournaud}, {Brown}, {Caputi}, {Casertano}, {Cassata}, {Castellano},
  {Challis}, {Chary}, {Cheung}, {Cirasuolo}, {Conselice}, {Roshan Cooray},
  {Croton}, {Daddi}, {Dahlen}, {Dav{\'e}}, {de Mello}, {Dekel}, {Dickinson},
  {Dolch}, {Donley}, {Dunlop}, {Dutton}, {Elbaz}, {Fazio}, {Filippenko},
  {Finkelstein}, {Fontana}, {Gardner}, {Garnavich}, {Gawiser}, {Giavalisco},
  {Grazian}, {Guo}, {Hathi}, {H{\"a}ussler}, {Hopkins}, {Huang}, {Huang},
  {Jha}, {Kartaltepe}, {Kirshner}, {Koo}, {Lai}, {Lee}, {Li}, {Lotz}, {Lucas},
  {Madau}, {McCarthy}, {McGrath}, {McIntosh}, {McLure}, {Mobasher},
  {Moustakas}, {Mozena}, {Nandra}, {Newman}, {Niemi}, {Noeske}, {Papovich},
  {Pentericci}, {Pope}, {Primack}, {Rajan}, {Ravindranath}, {Reddy}, {Renzini},
  {Rix}, {Robaina}, {Rodney}, {Rosario}, {Rosati}, {Salimbeni}, {Scarlata},
  {Siana}, {Simard}, {Smidt}, {Somerville}, {Spinrad}, {Straughn}, {Strolger},
  {Telford}, {Teplitz}, {Trump}, {van der Wel}, {Villforth}, {Wechsler},
  {Weiner}, {Wiklind}, {Wild}, {Wilson}, {Wuyts}, {Yan}, \& {Yun}}]{grogin+11}
{Grogin}, N.~A., {Kocevski}, D.~D., {Faber}, S.~M., {et~al.} 2011, \apjs, 197,
  35, 35

\bibitem[{{Inoue} {et~al.}(2014){Inoue}, {Shimizu}, {Iwata}, \&
  {Tanaka}}]{inoue+14}
{Inoue}, A.~K., {Shimizu}, I., {Iwata}, I., \& {Tanaka}, M. 2014, \mnras, 442,
  1805, 1805

\bibitem[{{Izotov} {et~al.}(2016){Izotov}, {Orlitov\'{a}}, {Schaerer}, {Thuan},
  {Verhamme}, {Guseva}, \& {Worseck}}]{izotov+16}
{Izotov}, Y.~I., {Orlitov\'{a}}, I., {Schaerer}, D., {et~al.} 2016, \nat, 529,
  159Ð160, 159Ð160

\bibitem[{{Izotov} {et~al.}(2014){Izotov}, {Thuan}, \& {Guseva}}]{izotov+14}
{Izotov}, Y.~I., {Thuan}, T.~X., \& {Guseva}, N.~G. 2014, \mnras, 445, 778, 778

\bibitem[{{James} {et~al.}(2016){James}, {Auger}, {Aloisi}, {Calzetti}, \&
  {Kewley}}]{james+16}
{James}, B.~L., {Auger}, M., {Aloisi}, A., {Calzetti}, D., \& {Kewley}, L.
  2016, \apj, 816, 40, 40

\bibitem[{{Jaskot} \& {Oey}(2013)}]{jaskotoey2013}
{Jaskot}, A.~E., \& {Oey}, M.~S. 2013, \apj, 766, 91, 91

\bibitem[{{Leitherer} {et~al.}(1999){Leitherer}, {Schaerer}, {Goldader},
  {Gonz{\'a}lez Delgado}, {Robert}, {Kune}, {de Mello}, {Devost}, \&
  {Heckman}}]{leitherer+99}
{Leitherer}, C., {Schaerer}, D., {Goldader}, J.~D., {et~al.} 1999, \apjs, 123,
  3, 3

\bibitem[{{Lusso} {et~al.}(2012){Lusso}, {Comastri}, {Simmons}, {Mignoli},
  {Zamorani}, {Vignali}, {Brusa}, {Shankar}, {Lutz}, {Trump}, {Maiolino},
  {Gilli}, {Bolzonella}, {Puccetti}, {Salvato}, {Impey}, {Civano}, {Elvis},
  {Mainieri}, {Silverman}, {Koekemoer}, {Bongiorno}, {Merloni}, {Berta}, {Le
  Floc'h}, {Magnelli}, {Pozzi}, \& {Riguccini}}]{lusso+12}
{Lusso}, E., {Comastri}, A., {Simmons}, B.~D., {et~al.} 2012, \mnras, 425, 623,
  623

\bibitem[{{Maseda} {et~al.}(2014){Maseda}, {van der Wel}, {Rix}, {da Cunha},
  {Pacifici}, {Momcheva}, {Brammer}, {Meidt}, {Franx}, {van Dokkum},
  {Fumagalli}, {Bell}, {Ferguson}, {F{\"o}rster-Schreiber}, {Koekemoer}, {Koo},
  {Lundgren}, {Marchesini}, {Nelson}, {Patel}, {Skelton}, {Straughn}, {Trump},
  \& {Whitaker}}]{maseda+14}
{Maseda}, M.~V., {van der Wel}, A., {Rix}, H.-W., {et~al.} 2014, \apj, 791, 17,
  17

\bibitem[{{Mostardi} {et~al.}(2013){Mostardi}, {Shapley}, {Nestor}, {Steidel},
  {Reddy}, \& {Trainor}}]{mostardi+13}
{Mostardi}, R.~E., {Shapley}, A.~E., {Nestor}, D.~B., {et~al.} 2013, \apj, 779,
  65, 65

\bibitem[{{Mostardi} {et~al.}(2015){Mostardi}, {Shapley}, {Steidel}, {Trainor},
  {Reddy}, \& {Siana}}]{mostardi+15}
{Mostardi}, R.~E., {Shapley}, A.~E., {Steidel}, C.~C., {et~al.} 2015, \apj,
  810, 107, 107

\bibitem[{{Nakajima} \& {Ouchi}(2014)}]{nakajimaouchi2014}
{Nakajima}, K., \& {Ouchi}, M. 2014, \mnras, 442, 900, 900

\bibitem[{{Nestor} {et~al.}(2013){Nestor}, {Shapley}, {Kornei}, {Steidel}, \&
  {Siana}}]{nestor+13}
{Nestor}, D.~B., {Shapley}, A.~E., {Kornei}, K.~A., {Steidel}, C.~C., \&
  {Siana}, B. 2013, \apj, 765, 47, 47

\bibitem[{{Oesch} {et~al.}(2015){Oesch}, {van Dokkum}, {Illingworth},
  {Bouwens}, {Momcheva}, {Holden}, {Roberts-Borsani}, {Smit}, {Franx},
  {Labb{\'e}}, {Gonz{\'a}lez}, \& {Magee}}]{oesch+15}
{Oesch}, P.~A., {van Dokkum}, P.~G., {Illingworth}, G.~D., {et~al.} 2015,
  \apjl, 804, L30, L30

\bibitem[{{Panessa} {et~al.}(2006){Panessa}, {Bassani}, {Cappi}, {Dadina},
  {Barcons}, {Carrera}, {Ho}, \& {Iwasawa}}]{panessa+06}
{Panessa}, F., {Bassani}, L., {Cappi}, M., {et~al.} 2006, \aap, 455, 173, 173

\bibitem[{{Peng} {et~al.}(2010){Peng}, {Ho}, {Impey}, \& {Rix}}]{peng+10}
{Peng}, C.~Y., {Ho}, L.~C., {Impey}, C.~D., \& {Rix}, H.-W. 2010, \aj, 139,
  2097, 2097

\bibitem[{{Prochaska} {et~al.}(2010){Prochaska}, {O'Meara}, \&
  {Worseck}}]{prochaska+10}
{Prochaska}, J.~X., {O'Meara}, J.~M., \& {Worseck}, G. 2010, \apj, 718, 392,
  392

\bibitem[{{Rafelski} {et~al.}(2015){Rafelski}, {Teplitz}, {Gardner}, {Coe},
  {Bond}, {Koekemoer}, {Grogin}, {Kurczynski}, {McGrath}, {Bourque}, {Atek},
  {Brown}, {Colbert}, {Codoreanu}, {Ferguson}, {Finkelstein}, {Gawiser},
  {Giavalisco}, {Gronwall}, {Hanish}, {Lee}, {Mehta}, {de Mello},
  {Ravindranath}, {Ryan}, {Scarlata}, {Siana}, {Soto}, \&
  {Voyer}}]{rafelski+15}
{Rafelski}, M., {Teplitz}, H.~I., {Gardner}, J.~P., {et~al.} 2015, \aj, 150,
  31, 31

\bibitem[{{Schlafly} \& {Finkbeiner}(2011)}]{schlaflyfinkbeiner11}
{Schlafly}, E.~F., \& {Finkbeiner}, D.~P. 2011, \apj, 737, 103, 103

\bibitem[Siana et al. (2007)]{siana+07}
Siana, B., Teplitz, H., Colbert, J., et al., 2007, \apj, 668, 62

\bibitem[{{Siana} {et~al.}(2010){Siana}, {Teplitz}, {Ferguson}, {Brown},
  {Giavalisco}, {Dickinson}, {Chary}, {de Mello}, {Conselice}, {Bridge},
  {Gardner}, {Colbert}, \& {Scarlata}}]{siana+10}
{Siana}, B., {Teplitz}, H.~I., {Ferguson}, H.~C., {et~al.} 2010, \apj, 723,
  241, 241

\bibitem[{{Siana} {et~al.}(2015){Siana}, {Shapley}, {Kulas}, {Nestor},
  {Steidel}, {Teplitz}, {Alavi}, {Brown}, {Conselice}, {Ferguson}, {Dickinson},
  {Giavalisco}, {Colbert}, {Bridge}, {Gardner}, \& {de Mello}}]{siana+15}
{Siana}, B., {Shapley}, A.~E., {Kulas}, K.~R., {et~al.} 2015, \apj, 804, 17, 17

\bibitem[{{Skelton} {et~al.}(2014){Skelton}, {Whitaker}, {Momcheva}, {Brammer},
  {van Dokkum}, {Labb{\'e}}, {Franx}, {van der Wel}, {Bezanson}, {Da Cunha},
  {Fumagalli}, {F{\"o}rster Schreiber}, {Kriek}, {Leja}, {Lundgren}, {Magee},
  {Marchesini}, {Maseda}, {Nelson}, {Oesch}, {Pacifici}, {Patel}, {Price},
  {Rix}, {Tal}, {Wake}, \& {Wuyts}}]{skelton+14}
{Skelton}, R.~E., {Whitaker}, K.~E., {Momcheva}, I.~G., {et~al.} 2014, \apjs,
  214, 24, 24

\bibitem[{{Stark} {et~al.}(2014){Stark}, {Richard}, {Siana}, {Charlot},
  {Freeman}, {Gutkin}, {Wofford}, {Robertson}, {Amanullah}, {Watson}, \&
  {Milvang-Jensen}}]{stark+14b}
{Stark}, D.~P., {Richard}, J., {Siana}, B., {et~al.} 2014, \mnras, 445, 3200,
  3200

\bibitem[{{Stern}(2015)}]{stern+15}
{Stern}, D. 2015, \apj, 807, 129, 129

\bibitem[{{Teplitz} {et~al.}(2013){Teplitz}, {Rafelski}, {Kurczynski}, {Bond},
  {Grogin}, {Koekemoer}, {Atek}, {Brown}, {Coe}, {Colbert}, {Ferguson},
  {Finkelstein}, {Gardner}, {Gawiser}, {Giavalisco}, {Gronwall}, {Hanish},
  {Lee}, {de Mello}, {Ravindranath}, {Ryan}, {Siana}, {Scarlata}, {Soto},
  {Voyer}, \& {Wolfe}}]{teplitz+13}
{Teplitz}, H.~I., {Rafelski}, M., {Kurczynski}, P., {et~al.} 2013, \aj, 146,
  159, 159

\bibitem[{{Ueda} {et~al.}(2015){Ueda}, {Hashimoto}, {Ichikawa}, {Ishino},
  {Kniazev}, {V{\"a}is{\"a}nen}, {Ricci}, {Berney}, {Gandhi}, {Koss},
  {Mushotzky}, {Terashima}, {Trakhtenbrot}, \& {Crenshaw}}]{ueda+15}
{Ueda}, Y., {Hashimoto}, Y., {Ichikawa}, K., {et~al.} 2015, \apj, 815, 1, 1

\bibitem[{{Vanzella} {et~al.}(2010{\natexlab{a}}){Vanzella}, {Siana},
  {Cristiani}, \& {Nonino}}]{vanzella+10b}
{Vanzella}, E., {Siana}, B., {Cristiani}, S., \& {Nonino}, M.
  2010{\natexlab{a}}, \mnras, 404, 1672, 1672

\bibitem[{{Vanzella} {et~al.}(2010{\natexlab{b}}){Vanzella}, {Giavalisco},
  {Inoue}, {Nonino}, {Fontanot}, {Cristiani}, {Grazian}, {Dickinson}, {Stern},
  {Tozzi}, {Giallongo}, {Ferguson}, {Spinrad}, {Boutsia}, {Fontana}, {Rosati},
  \& {Pentericci}}]{vanzella+10c}
{Vanzella}, E., {Giavalisco}, M., {Inoue}, A.~K., {et~al.} 2010{\natexlab{b}},
  \apj, 725, 1011, 1011

\bibitem[{{Vanzella} {et~al.}(2012){Vanzella}, {Guo}, {Giavalisco}, {Grazian},
  {Castellano}, {Cristiani}, {Dickinson}, {Fontana}, {Nonino}, {Giallongo},
  {Pentericci}, {Galametz}, {Faber}, {Ferguson}, {Grogin}, {Koekemoer},
  {Newman}, \& {Siana}}]{vanzella+12}
{Vanzella}, E., {Guo}, Y., {Giavalisco}, M., {et~al.} 2012, \apj, 751, 70, 70

\bibitem[{{Vanzella} {et~al.}(2015){Vanzella}, {de Barros}, {Castellano},
  {Grazian}, {Inoue}, {Schaerer}, {Guaita}, {Zamorani}, {Giavalisco}, {Siana},
  {Pentericci}, {Giallongo}, {Fontana}, \& {Vignali}}]{vanzella+15}
{Vanzella}, E., {de Barros}, S., {Castellano}, M., {et~al.} 2015, \aap, 576,
  A116, A116

\bibitem[{{Verhamme} {et~al.}(2015){Verhamme}, {Orlitov{\'a}}, {Schaerer}, \&
  {Hayes}}]{verhamme+15}
{Verhamme}, A., {Orlitov{\'a}}, I., {Schaerer}, D., \& {Hayes}, M. 2015, \aap,
  578, A7, A7


\bibitem[{{Wise} {et~al.}(2014){Wise}, {Demchenko}, {Halicek}, {Norman},
  {Turk}, {Abel}, \& {Smith}}]{wise+14}
{Wise}, J.~H., {Demchenko}, V.~G., {Halicek}, M.~T., {et~al.} 2014, \mnras,
  442, 2560, 2560

\bibitem[{{Zitrin} {et~al.}(2015){Zitrin}, {Labb{\'e}}, {Belli}, {Bouwens},
  {Ellis}, {Roberts-Borsani}, {Stark}, {Oesch}, \& {Smit}}]{zitrin+15}
{Zitrin}, A., {Labb{\'e}}, I., {Belli}, S., {et~al.} 2015, \apjl, 810, L12, L12

\end{thebibliography}

\end{document}